\documentclass{moriond_Arxiv}

\def\Journal#1#2#3#4{{#1} {\bf #2}, #3 (#4)}

\usepackage{authblk}

\title{Status of the COSmological Microwave Observations CALibrator}

\author[1]{L.~Mousset}
\author[2]{J.~Aumont}
\author[3]{S.~Berta}
\author[4]{L.~Bizzarri}
\author[5]{F.~Boulanger}
\author[6]{A.~Catalano}
\author[7]{F.~Cuttaia}    
\author[5]{A.~Denis}
\author[3]{A.-L.~Fontana}
\author[8]{D.~Gonz\'alez Ovejero}
\author[6]{B.~Kalemi}   
\author[3]{S.~Leclercq}
\author[6]{J.-F.~Mac\'ias-P\'erez}
\author[1]{B.~Maffei}
\author[10]{T.~Mcnamara}
\author[6]{S.~Micheli}
\author[9]{M.~Migliaccio}
\author[2]{L.~Montier}    
\author[5]{P.~Morfin}
\author[10]{M.~Moulin}
\author[11]{M.~Murgia}
\author[12]{I.~Myserlis}
\author[4]{F.~Nati}
\author[11]{P.~Ortu}
\author[5]{M.~Pérault}
\author[13]{G.~Pisano}    
\author[11]{T.~Pisanu}
\author[14]{N.~Ponthieu}
\author[6]{A.~Ritacco}
\author[15]{S.~Savorgnano}
\author[7]{L.~Terenzi}
\author[16]{L.~Thibaut}    
\author[17]{J.~Treuttel}
\author[6]{C.~Vescovi}
\author[16]{L.~Vacher}
\author[4]{M.~Zannoni}

\affil[1]{Institut d’Astrophysique Spatiale, CNRS-Université Paris-Saclay, France}
\affil[2]{Institut de Recherche en Astrophysique et Planétologie, CNRS-INSU, France}
\affil[3]{Institut de Radio-Astronomie Millimétrique, France}
\affil[4]{Department of Physics, University of Milano-Bicocca, Italy}
\affil[5]{Laboratoire de Physique de l’École normale supérieure, ENS, Université PSL, CNRS, Sorbonne Université, Université Paris Cité, F-75005 Paris, France}
\affil[6]{Laboratoire de Physique Subatomique et de Cosmologie, CNRS, France}
\affil[7]{INAF - Osservatorio di astrofisica e scienza dello spazio Bologna, Italy}
\affil[8]{Institut d'Electronique et des Technologies du numérique, France}
\affil[9]{University of Rome Tor Vergata, Italy}
\affil[10]{Centre spatial universitaire de Grenoble, Université Grenoble Alpes, France}
\affil[11]{INAF - Osservatorio Astronomico di Cagliari, Italy}
\affil[12]{Institut de Radioastronomie Millim\'{e}trique, Avenida Divina Pastora, 7, Local 20, E--18012 Granada, Spain}
\affil[13]{Sapienza Università di Roma, Italy}
\affil[14]{Institut de Planétologie et d'Astrophysique de Grenoble, CNRS, France}
\affil[15]{Department of Physics, Boston University, United States}
\affil[16]{Laboratoire de Physique des 2 infinis Irène Joliot-Curie, CNRS, France}
\affil[17]{LERMA, Observatoire de Paris-PSL, CNRS, France}

\date{}

\begin{document}

\maketitle\abstracts{As the sensitivity of CMB telescopes increases, the need for precise calibration becomes critical. Started in 2022, the COSMOCal project aims to place an artificial polarized source in geostationary orbit, which will serve as a reference for CMB telescopes. This source will emit at 90, 150 and 270 GHz and will be linearly polarized with a highly precise orientation smaller than $0.1\deg$. This proceeding presents the scientific motivations for the project, the current status of the development of the instrument and the results of a calibration campaign performed in March 2026 at the Institut d’Astrophysique Spatiale.}

\section{Introduction}
The Cosmic Microwave Background (CMB) is a powerful probe of the early universe and has been extensively studied over the past few decades. The measurement of CMB polarization, in particular, has opened new windows into the physics of the early universe, such as the search for primordial gravitational waves and the characterization of astrophysical foregrounds. However, the measurement of CMB polarization is challenging and requires precise calibration of the instruments. 

As instruments become more and more sensitive, the susceptibility to systematic effects increases and so the requirements on calibration become more stringent. In particular, the calibration of the beam and the polarization angle of CMB telescopes is becoming a major issue for the next generation of experiments, such as the Simons Observatory (SO)~\cite{SO}, and there is a clear need for new calibration methods.

The COSMOCal project aims to address this issue by providing an artificial polarized source in geostationary orbit, which will serve as a reference for CMB telescopes~\cite{Ritacco}. The source will emit at 90, 150 and 270 GHz and will be linearly polarized with an angle known in the celestial reference frame with an accuracy of 0.1°. It will be visible from Chile and Europe. This goes in the same direction as the POLOCALC project which aims to provide a calibration source on a drone platform for CMB telescopes~\cite{Nati}.

\section{Scientific motivations}\label{sec:motiv}

\subsection{Cosmic birefringence}
A first strong motivation for a precise calibration of the polarization angle is the search for cosmic birefringence. As CMB photons propagate through the universe, they can couple with pseudo-scalar field (e.g. axions) causing their linear polarization to rotate. This can cause E to B~mode leakage, leading to a non-zero EB correlation. 

Today, both Atacama Cosmology Telescope (ACT) and Planck see a slight preference for non zero cosmic birefringence~\cite{Diego-Palazuelos_Planck,Diego-Palazuelos_ACT}. With Planck data release 4 they find a cosmic birefringence angle of $\beta = 0.30 \pm 0.11\deg$ and with ACT data, $\beta = 0.215 \pm 0.074\deg$. So, if we were to combine the results we would get a 4~sigma detection. However, without absolute angle calibration, we will not know whether both results are affected by miscalibrated polarization angles. Thus, an absolute calibration of the polarization angle with an accuracy of $0.1\deg$ is required to confirm the detection of cosmic birefringence.

\subsection{A need for beam calibration in polarization}

A motivation for beam calibration comes from recent results from the ACT and the South Pole Telescope (SPT) collaborations. Both experiments have measured a deviation from the Planck EE power spectrum at high multipoles. This could be a hint for new physics but there is a risk that it is due to a miscalibration of the polarized beams.

In intensity, the beam of a telescope can be measured using planets, which are bright and unpolarized sources. However, in polarization, there is no such astrophysical source that would be known with the required precision. Thus, an artificial source in geostationary orbit would provide a stable and well-characterized source for beam calibration in polarization.

\subsection{Astrophysical foregrounds characterization}

A major source of contamination for CMB polarization measurement is the presence of astrophysical foregrounds, such as galactic dust and synchrotron emission. These foregrounds are polarized and can mimic the CMB polarization signal. We know that the foreground polarization is complex, with variations of the polarization angle across frequencies and across the sky. This complexity can lead to E to B leakage, which can contaminate the BB and EB power spectra~\cite{Ritacco_dust,Vacher,Guillet}.

Thus, calibration of the polarization angle measurement would allow us to better characterize the foreground emissions and to better understand their complexity, which is crucial for the component separation step.

\section{The COSMOCal project}\label{sec:project}

\subsection{COSMOCal as a reference for CMB telescopes}

To be in the far field of large aperture telescopes, it is necessary to place the source in space. For example, for the the SO-Large Aperture Telescope (SO-LAT), which has a diameter of 6~m, the far field starts at around 36~km at 150\,GHz.~\footnote{The far field distance $d$ of an antenna is given by $d=2 D^2/\lambda$ with $D$ the diameter of the antenna and $\lambda$ the wavelength.}

The COSMOCal source will be visible from few telescopes in Chile and in Europe: the SO-LAT, the 30\,m IRAM telescope and the Sardinia Radio Telescope (SRT). Those instruments will be able to directly observe the source and perform beam calibration and absolute calibration of the polarization angle. 

Once those instruments will be calibrated, they will provide reference CMB polarization maps so that the other telescopes, such as LiteBIRD~\cite{LiteBIRD}, will be able to cross calibrate their measurements. In particular, the SO-LAT could provide a measurement, or at least an upper limit, of the cosmic birefringence angle $\beta$, so that the other telescopes can calibrate their polarisation angle using the EB cross-spectra.

\subsection{Current status of the project}

The project started in 2022 and is currently in the development phase, monitored by CNES. The COSMOCal instrument is designed to be a guest payload on a satellite placed in geostationary orbit and operated by the private company Eutelsat. The expected CNES Phase A review is scheduled for early 2027, which will validate (or not) the launch with Eutelsat in 2030. Another option, for a later launch, is also under study in case the one with Eutelsat would be unfeasible.

\subsection{Design of the instrument}

A first prototype at 260~GHz was fully assembled at LPENS in Paris in 2023~\cite{Ritacco}. It was then characterized at LPSC in Grenoble in February 2024 and a calibration campaign was performed at the 30\,m telescope of IRAM in Spain in September 2024.

\begin{figure}[ht!]
\centering
\includegraphics[width=0.4\textwidth]{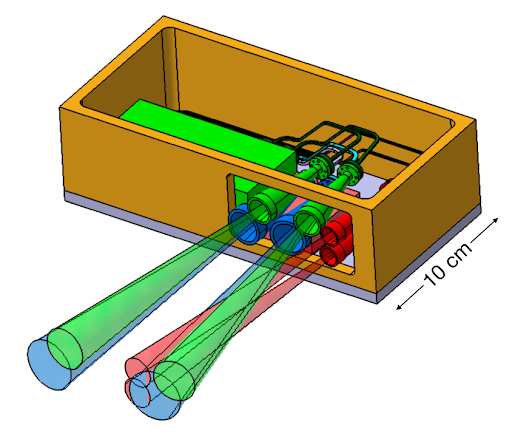}
\caption{Schematic design of the COSMOCal instrument with horns at 90, 150 and 270 GHz emitting to Chile and Europe. Credit : P. Morfin.}
\label{fig:newdesign}
\end{figure}

Since that time, the design has been considerably simplified and the new design is under study. As shown in Figure~\ref{fig:newdesign}, a possibility is to have three emitting frequencies at 90, 150 and 270 GHz, with horns pointing to Chile (SO-LAT) and Europe. The instrument orientation would then be controlled by the altitude control of the satellite and the polarization angle would be calibrated using pre-launch laboratory measurements. 

\subsection{Calibration campaign at IAS}
\begin{figure}[ht!]
\centering
\includegraphics[width=0.45\textwidth]{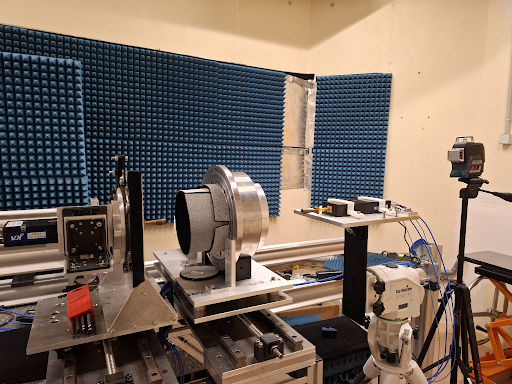}
\includegraphics[width=0.5\textwidth]{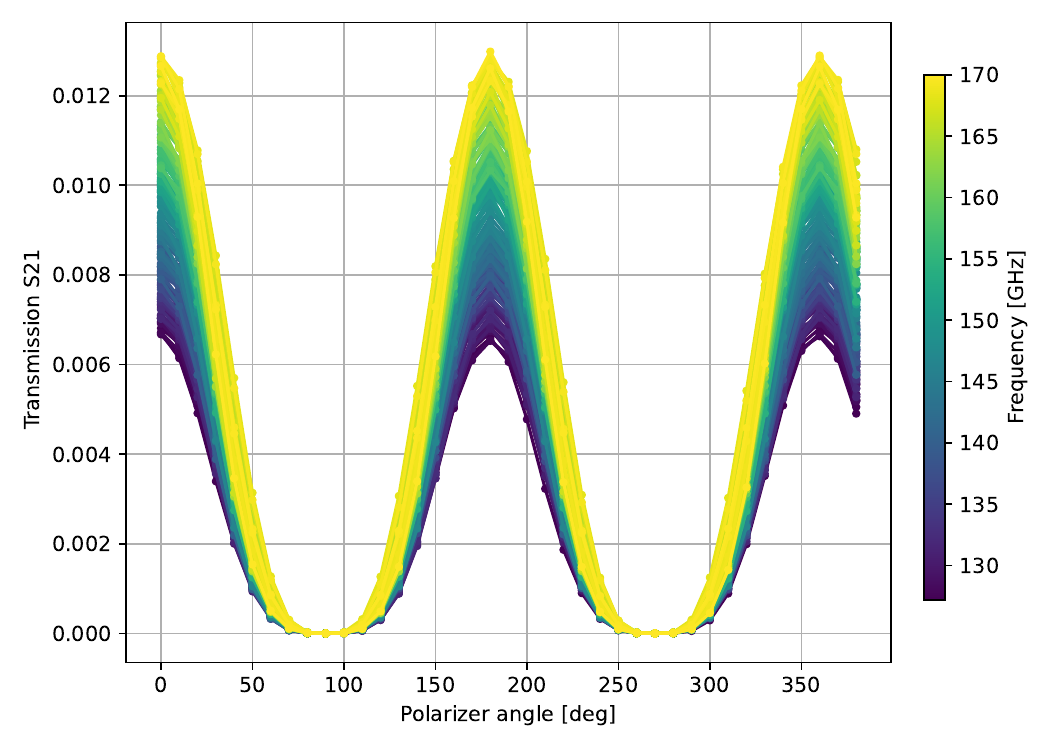}
\caption{Calibration campaign at IAS. \textit{Left:} The picture shows the two horns (emitter on the right and receiver on the left) and a polarizer in between, placed on a rotating stage covered with absorbing material.
\textit{Right:} Transmission coefficient $S_{21}$ measured by the VNA as a function of the polarizer angle for each frequency sample.}
\label{fig:IAS}
\end{figure}

In March 2026, at IAS in Orsay, a calibration campaign was conducted to establish a method to measure the polarization angle of the instrument with an accuracy of $0.1\deg$. This measurement must be done with respect to a reference frame attached to the instrument. For that purpose, we used two horns, an emitter that plays the role of the COSMOCal instrument, and a receiver, both connected to a Vector Network Analyzer (VNA). A picture of the setup is shown in Figure~\ref{fig:IAS} (left). The VNA was scanning the frequency range between 110 and 170~GHz with 242 data points linearly spaced in frequency. The receiver was placed on a rotating stage that we could rotate manually with a precision of $0.5\deg$ approximately. We also added a polarizer between the two horns, placed on a motorized rotating mount. By using a microscope, we were able to align the polarizer wires with respect to its metallic mount with an accuracy of few arcminutes. Then, the wires were placed horizontally, using a bubble level placed on the polarizer mount. By doing so, we defined a reference frame for the polarization angle measurement. We then measured the power received by the horn as a function of the polarizer orientation. In Figure~\ref{fig:IAS} (right), we show the transmission coefficient $S_{21}$ measured by the VNA as a function of the polarizer angle for each frequency sample. In this measurement, the two horns were aligned so that $S_{21}$ follows a cosine power four law. The differences in amplitude between the curves can be explained by the increase in the gain of the horns as frequency increases. By repeating this measurement with different orientations between the two horns and by fitting the curves, we should be able to determine the instrument polarization angle of the instrument. Unfortunately, due to systematics, probably caused by stationary waves and reflections in the setup, we were not able to reach the required accuracy. We are currently working on improving the setup in order to reach this goal.

\section{Conclusion}

The calibration of polarization measurements from CMB telescopes is a key issue, one that becomes even more important as the sensitivity of the instruments improves. COSMOCal is part of a broader effort by the CMB community to improve instrument calibration. The goal is to be ready for the launch in 2030 with the Eutelsat company, which is ambitious but also timely with the SO calendar.

\section*{Acknowledgments}
The authors acknowledge financial support from the Centre national d’études spatiales (CNES), France (ROR: https://ror.org/04h1h0y33), within the framework of the COSMOCal space mission. F. Nati acknowledges funding from the European Union (ERC, POLOCALC, 101096035). 

\section*{References}

\begin{thebibliography}{99}

\bibitem{SO}Simons Observatory collaboration, \Journal{JCAP}{02}{056}{2019}.
\bibitem{Ritacco}A. Ritacco {\it et al}, \Journal{Publ. Astron. Soc. Pac.}{136}{11}{2024}.
\bibitem{Nati}F. Nati {\it et al}, \Journal{Journal of Astronomical Instrumentation}{6}{1740008}{2017}.
\bibitem{Diego-Palazuelos_Planck}P. Diego-Palazuelos, and others, \Journal{Phys. Rev. Lett.}{128}{091302}{2022}.
\bibitem{Diego-Palazuelos_ACT}P. Diego-Palazuelos and E. Komatsu, \Journal{arXiv:2509.13654}{2509}{13654}{2025}.
\bibitem{Ritacco_dust}A. Ritacco {\it et al}, \Journal{A\&A}{670}{A163}{2023}.
\bibitem{Vacher}L. Vacher {\it et al}, \Journal{A\&A}{672}{A146}{2023}.
\bibitem{Guillet}V. Guillet {\it et al}, \Journal{A\&A}{705}{A177}{2026}.
\bibitem{LiteBIRD}LiteBIRD collaboration, \Journal{Progress of Theoretical and Experimental Physics}{2023}{4}{2023}.

\end{thebibliography}


\end{document}